\documentclass[aps,twocolumn,pra,superscriptaddress,showpacs,tightenlines]{revtex4}
\usepackage{amssymb}
\usepackage{amsmath}
\usepackage{graphicx}
\usepackage{epsfig}
\usepackage{subfigure}

\begin{document}

\title{Decoherence problem in quantum state transfer via an engineered spin
chain}
\author{Lan \surname{Zhou}}
\affiliation{Institute of Theoretical Physics, Chinese Academy of Sciences, Beijing,
100080,China}
\author{Jing \surname{Lu}}
\affiliation{Department of Physics, Hunan Normal University, Changsha 410081, China}
\author{Tao \surname{Shi}}
\affiliation{Department of Physics, Nankai University, Tianjin 300071, China}
\author{C. P. \surname{Sun}}
\email{suncp@itp.ac.cn} \homepage{http://www.itp.ac.cn/~suncp}
\affiliation{Institute of Theoretical Physics, Chinese Academy of Sciences, Beijing,
100080,China}

\begin{abstract}
A perfect quantum state transfer(QST) has been shown in an
engineered spin chain with ``always-on interaction". Here, we
consider a more realistic problem for such a protocol, the quantum
decoherence induced by a spatially distributed environment, which is
universally modeled as a bath of harmonic oscillators. By making use
of the irreducible tensor method in angular momentum theory, we
investigate the effect of decoherence on the efficiency of QST for
both cases at zero and finite temperatures. We not only show the
generic exponential decay of QST efficiency as the number of sites
increase, but also find some counterintuitive effect, the QST can be
enhanced as temperature increase.
\end{abstract}

\maketitle

\section{Introduction}

In quantum communications, one of the most important tasks is
transmitting a quantum state (known or unknown) from one location to
another location. The ideal channel for accomplishing this task is
to share entanglement with a separated party for
teleportation\cite{Bennett}, but measurement is required to obtain
the original quantum state. Most recently, people has explored the
new possibility to transmit a quantum state through a data bus with
``always-on" interaction, where minor operations are required. Among
such quantum state transfer (QST) schemes, the protocols based on
the spin chain with engineered couplings are particular attractive
for short distance communication\cite{Bose1,Plenio1,Chris1,CPSun1},
as it does not require any gating, modulating and measurements.

However a real quantum system can rarely be isolated from its surrounding
environment completely. It is usually coupled to the external environment
(also called " bath") with a large number of degrees of freedom\cite%
{CPSun2}. There are two distinct effects of the bath on the quantum
system,  quantum dissipation and decoherence. The difference between
dissipation and decoherence is whether coupling between system and
bath allows for an exchange of energy \cite{CPSun2}. The dissipation
effect of a spin bath on spin transfer functions of permanently
coupled spin system have been investigated in Ref. \cite{Bose2}, and
the decoherence effects of the efficiency of the spin chain have
also been studied in Ref.\cite{ZhouGuo}. In Ref. \cite{Bose2} an
assumption was made, that is, each spin of the spin chain was
assumed to couple independently to separate baths. It is an ideal
case. It is more practical to study the components of the system
interacting with the same environment.

In this paper, we consider a system interacts with a spatially
distributed globe bath by exchange interactions. Here, the system is
an artificial chain with engineered nearest neighbor couplings and
the bath is generally treated as a collection of two-level state
systems. Through irreducible tensor method in angular momentum
theory, we use the Langevin approach\cite{luisell} to study the
environmental effects on the efficiency of this QST system. We find
that the transport efficiency decays exponentially with time; when
the norm of the amplitudes of the transport state are equal,
transport efficiency is independent of temperature. But when the
norm of the amplitudes are not equal, the transport efficiency is a
decreasing or a increasing function of temperature respectively; it
also decreases as the number of sites increased.

This paper is organized as follows: In section \ref{sec:model}, we present
our model for the coupling between the system and the bath. In section \ref%
{sec:Langevin}, we derive the time evolution of system with irreducible
tensor method and the Langevin approach. In section \ref{sec:effect}, we
investigate the the bath effect on the state transport in the first
excitation subspace. In section \ref{sec:temper}, we investigate the effect
on state transferring at finite temperature. In Section \ref{sec:end}, we
make our conclusion.

\section{\label{sec:model}model}

The decoherence model we consider here is a N-sites lattice
interacts with a heat bath by exchange interactions. The system for
QST is an open chain with engineered nearest neighbor couplings
proposed in Ref.\cite{Chris1}. The Hamiltonian for the spinless
fermion in a tight binding lattice is
\begin{equation}
H_{s}=\frac{\theta }{2}\sum_{l=1}^{N-1}J_{l}(\hat{a}_{l}^{\dagger }\hat{a}
_{l+1}+\hat{a}_{l+1}^{\dagger }\hat{a}_{l}).
\label{Hsystem}
\end{equation}
It describes the  free hopping in a network of N
lattice sites, which is  equivalent to the XY-spin model. Here, $\hat{a}%
_{l}^{\dagger }$($\hat{a}_{l}$) is the fermion creation(annihilation)
operator of electrons at $l$th site; $J_{l}=\sqrt{l(N-l)}$ is the hopping
integral over the $l$th site and the $l+1$th site. In the single fermion
subspace of the lattice, $H_{s}=\theta \hat{J}_{x}$, where $\hat{J}_{x}$ is
the angular momentum operator for a particle of spin $S=(N-1)/2$, scaled by
a strength parameter $\theta $. The single-fermion states $\{|l\rangle
=a_{l}^{\dagger }|vac\rangle \}$ are equivalent to the z-angular momentum
eigenstates, with $\{|1\rangle =|J,-J\rangle ,\cdots ,|N\rangle =|J,J\rangle
\}$. Then through the time evolution driven by $H_{s}$, a perfect QST is
achieved at time $t=\pi /\theta $.

The  bath around the open chain distributes as a large ensemble of
two-level systems at different position without interaction. Each
position is represented by a site. The state of site is expressed as
being empty or occupied by a spinless fermionic particle. This
modeling of the bath can be realized as the background charge
fluctuation for the practical electron system. The Hamiltonian of
the distributed thermal bath has the following form
\begin{equation}
H_{B}=\sum_{x=1}^{M-1}\omega_{x}\hat{b}_{x}^{\dagger}\hat{b}_{x},
\label{Hbath}
\end{equation}
where $\omega_{x}$ is the on-site potential (or called the chemical
potential) of the $x$th site. $\hat{b}_{x}^{\dagger}$ ($\hat{b}_{x}$) stands
the fermion creation(annihilation) operator of electrons at $x$th site. And
$M$ is the number of sites. By the rotating-wave approximation, the
Hamiltonian for the exchange interaction between the system and the bath is
written as
\begin{equation}
H_{I}=\sum_{l,x=1}^{N,M}(g_{lx}\hat{a}_{l}^{\dagger}\hat{b}_{x}+h.c.),
\label{HI}
\end{equation}
where $g_{lx}$ is the coupling strength between the $l$th site of system and
the $x$th site of bath. Obviously, $H_{I}$ describes the system interacting
with a globe environment, which is more practical.

The total Hamiltonian of this entire system is described by
\begin{equation}
\tilde{H}=H_{s}+H_{B}+H_{I},  \label{Htide}
\end{equation}
which preserves the total number of excitations. It is important to
note that $[H_{s},H_{I}]\neq 0$, which means the system and the bath
exchange energy with each other. This is typical quantum dissipation
problem.

\section{\label{sec:Langevin}Quantum Langevin approach for system evolution}

Not to be limited by the concrete initial states of the system, the Langevin
approach is used to derive the time evolution of the system operator. First,
we employ the irreducible tensor method in angular momentum theory to
diagonalize Hamiltonian $H_{s}$. By defining a new fermion operators $\hat{c}
_{l}^{\dag }$, $\hat{c}_{l}$
\begin{subequations}
\begin{eqnarray}
\hat{c}_{l} &=&\sum_{l^{\prime }=1}^{N}d_{ll^{\prime }}\left( -\frac{\pi }{2}
\right) \hat{a}_{l^{\prime }},  \label{ccplus} \\
\hat{c}_{l}^{\dag } &=&\sum_{l^{\prime }=1}^{N}d_{l^{\prime }l}\left( \frac{
\pi }{2}\right) \hat{a}_{l^{\prime }}^{\dagger },
\end{eqnarray}
\end{subequations}
where
\begin{eqnarray}
d_{jl}\left( \frac{\pi }{2}\right) &=&2^{\frac{1-N}{2}}\sqrt{\left(
l-1\right) !\left( N-l\right) !\left( j-1\right) !\left( N-j\right) !}
\notag \\
&&\sum_{\nu }(-1)^{j-l+\nu }\left[ \left( N-j-\nu \right) !\left( l-1-\nu
\right) !\right] ^{-1}  \notag \\
&&\left[ \left( \nu +j-l\right) !\nu !\right] ^{-1}.  \label{dl'l}
\end{eqnarray}%
The total Hamiltonian $\tilde{H}$ can be written as
\begin{eqnarray}
H &=&\sum_{l=1}^{N}\theta l\hat{c}_{l}^{\dag }\hat{c}_{l}+\sum_{x=1}^{M}%
\omega _{x}\hat{b}_{x}^{\dag }\hat{b}_{x}  \notag \\
&&+\sum_{x=1}^{M}\sum_{l=1}^{N}\left( \tilde{g}_{lx}\hat{c}_{l}^{\dag }\hat{b%
}_{x}+h.c.\right),
\label{H}
\end{eqnarray}
where
\begin{equation}
\tilde{g}_{lx}=\sum_{j=1}^{N}g_{jx}d_{jl}\left( -\frac{\pi }{2}\right) .
\end{equation}
The system Hamiltonian (\ref{H}) described is analogous to a multilevel atom
coupling to a reservoir, with the atomic energy level spaced uniformly. It
is a typical damped system, and the interaction results in an atomic
linewidth.

The Heisenberg equation driven by the Hamiltonian (\ref{H}) results in the
following system of equations.
\begin{subequations}
\begin{eqnarray}
\frac{d\hat{c}_{m}^{\dagger }}{dt} &=&im\theta \hat{c}_{m}^{\dagger
}+i\sum_{x}^{M}\tilde{g}_{mx}^{\ast }\hat{b}_{x}^{\dag }, \\
\frac{d\hat{b}_{x}^{\dag }}{dt} &=&i\omega _{x}\hat{b}_{x}^{\dagger
}+i\sum_{l=1}^{N}\tilde{g}_{lx}\hat{c}_{l}^{\dagger }.
\end{eqnarray}
\label{eq-cb}
\end{subequations}
In general, Eq.(\ref{eq-cb}) cannot be solved exactly. We assume the
interaction between chain and bath is weak, and then can resort to
perturbation theory. First we define two new fermion operators $\hat{C}%
_{m}^{\dagger }$ and $\hat{B}_{x}^{\dag }$
\begin{subequations}
\begin{eqnarray}
\hat{c}_{m}^{\dagger } &=&\hat{C}_{m}^{\dagger }e^{i\theta mt} \\
\hat{b}_{x}^{\dagger } &=&\hat{B}_{x}^{\dag }e^{i\omega _{x}t}
\end{eqnarray}
\label{eq-ccbb}
\end{subequations}
to remove the high frequency effect, and then the Heisenberg equation for
operator $\hat{C}_{m}^{\dagger }$ and $\hat{B}_{x}^{\dag }$ becomes
\begin{subequations}
\begin{eqnarray}
\frac{d\hat{C}_{m}^{\dagger }}{dt} &=&i\sum_{x}^{M}\tilde{g}_{mx}^{\ast
}e_{x}^{-i\left( \theta m-\omega _{x}\right) \left( t-t_{0}\right) }\hat{B}%
_{x}^{\dag }, \\
\frac{d\hat{B}_{x}^{\dag }}{dt} &=&i\sum_{l=1}^{N}\tilde{g}_{lx}e^{i\left(
\theta l-\omega _{x}\right) \left( t-t_{0}\right) }\hat{C}_{l}^{\dagger }.
\end{eqnarray}
\label{eq-sb}
\end{subequations}
Integrating both sides of Eq.(\ref{eq-sb}), and iterating it up to second
order of $g$, we obtain the integral-differential equation
\begin{eqnarray}
\frac{d\hat{C}_{m}^{\dagger }}{dt} &=&i\sum_{x}^{M}\tilde{g}_{mx}^{\ast
}e^{-i\left( \theta m-\omega _{x}\right) t}\hat{B}_{x}^{\dag } \\
&&-\sum_{x,l=1}^{M,N}\tilde{g}_{mx}^{\ast }\tilde{g}_{lx}e^{-i\left( \theta
m-\omega _{x}\right) t}\hat{C}_{l}^{\dagger }\int_{0}^{t}dt^{\prime
}e^{i\left( \theta l-\omega _{x}\right) t^{\prime }}. \nonumber
\end{eqnarray}
Here, we assume the coupling between system and bath is started at $t_{0}=0$%
. By Laplace transformation, we have%
\begin{eqnarray}
&&\left[ s+\sum_{x=1}^{M}\frac{\tilde{g}_{mx}^{\ast }\tilde{g}_{mx}}{i\left(
\theta m-\omega _{x}\right) +s}\right] \hat{C}_{m}^{\dagger }\left( s\right) \nonumber \\
&=&\hat{C}_{m}^{\dagger }+i\sum_{x}^{M}\frac{\tilde{g}_{mx}^{\ast }}{i\left(
\theta m-\omega _{x}\right) +s}\hat{B}_{x}^{\dag } \nonumber \\
&&-\sum_{x=1}^{M}\sum_{l\neq m}^{N}\frac{\tilde{g}_{mx}^{\ast }\tilde{g}_{lx}
}{i\left( \theta m-\omega _{x}\right) +s}\frac{\hat{C}_{l}^{\dagger }}{
i\left( \theta m-\theta l\right) +s} .
\end{eqnarray}
Through Wigner-Weisskopf approximation\cite{luisell}, we obtain%
\begin{eqnarray}
\hat{C}_{m}^{\dagger }\left( t\right) &=&e^{-\Gamma _{m}t}\hat{C}%
_{m}^{\dagger }+i\sum_{x}^{M}\frac{\tilde{g}_{mx}^{\ast }e^{-\Gamma _{m}t}}{%
i\left( \theta m-\omega _{x}\right) -\Gamma _{m}}\hat{B}_{x}^{\dag }
\label{tt} \\
&&-\sum_{x=1}^{M}\sum_{l\neq m}^{N}\frac{\tilde{g}_{mx}^{\ast }\tilde{g}_{lx}%
}{i\left( \theta m-\omega _{x}\right) -\Gamma _{m}}\frac{\hat{C}%
_{l}^{\dagger }e^{-\Gamma _{m}t}}{i\theta \left( m-l\right) -\Gamma _{m}},
\notag
\end{eqnarray}
where the decay rate
\begin{equation}
\Gamma _{m}=\pi \sum_{x=1}^{M}\tilde{g}_{mx}^{\ast }\tilde{g}_{mx}\delta
\left( \omega _{x}-\theta m\right)  \label{decay}
\end{equation}
reflects the bath effect on the time evolution of the system.

\section{\label{sec:effect}effects on the state transport at zero temperature%
}

Now we consider the dissipation effect at zero temperature. Since the system
and the bath are independent before coupling, the initial state of the chain
and the bath is chosen to be factorized. As a state can be perfectly
transferred to the end of the chain after time $t=\pi /\theta $ in the first
excitation subspace, we assume the total system only has single fermionic
particle. The quantum state to be transferred is located at the $1$st site
of the chain, so the bath is initially in a vacuum state. We denote the
state of the chain as $|1\rangle =\hat{a}_{1}^{\dagger }|0\rangle $, where $%
|0\rangle $ is the vacuum state and it means there are no particles
in the chain. We then investigate how the bath affects the state
transferring from the $1$st site to the end of the chain. The
transport efficiency of the
chain is described by the norm of transfer function(MTF)\cite{Bose1,Chris1}%
\begin{equation}
F=tr\left[ U(t)|1\rangle \langle 1|\rho _{B}U^{\dagger }(t)|N\rangle \langle
N|\right] ,  \label{fizero}
\end{equation}
where $U(t)=\exp {(-i\tilde{H}t)}$ and $\rho _{B}=|0\rangle \langle 0|$ is
the initial state of the bath. The value of MTF is between 0 (no transfer)
and 1 (a perfect transfer). In terms of operators $\hat{C}^{\dagger }$ and $%
\hat{C}$, the probability for finding the particle in the $N$th site is
reexpressed as
\begin{eqnarray}
F &=&\sum_{mnll^{\prime }}d_{mN}\left( -\frac{\pi }{2}\right) d_{Nn}\left(
\frac{\pi }{2}\right) d_{1l}\left( \frac{\pi }{2}\right) d_{l^{\prime
}1}\left( -\frac{\pi }{2}\right) \nonumber \\
&&\left\langle 0\right\vert \hat{C}_{l}\left[ \hat{C}_{m}^{\dag }\left(
t\right) \hat{C}_{n}\left( t\right) \right] \hat{C}_{l^{\prime }}^{\dag
}\left\vert 0\right\rangle e^{i\theta \left( m-n\right) t}.
\end{eqnarray}
According to the discussions in the above section, and neglecting the cross
term $\tilde{g}_{nx}^{\ast }\tilde{g}_{lx}$ when $n\neq l$, we have
\begin{eqnarray}
F &=&\sum_{mn}d_{mN}\left( -\frac{\pi }{2}\right) d_{Nn}\left( \frac{\pi }{2}%
\right) d_{1m}\left( \frac{\pi }{2}\right) d_{n1}\left( -\frac{\pi }{2}%
\right) \nonumber \\
&&\exp \left[ i\theta \left( m-n\right) t-\left( \Gamma _{m}+\Gamma
_{n}\right) t\right].
\end{eqnarray}
If coupling constant $\tilde{g}_{lx}$ satisfying $\Gamma _{l}=\Gamma
$, the MTF become
\begin{equation}
F=(\sin \frac{\theta t}{2})^{2\left( N-1\right) }e^{-2\Gamma t},
\end{equation}
which shows that MTF decays exponentially with decay rate $2\Gamma $.

\begin{figure}[htbp]
\includegraphics[bb=109 348 354 746, width=7cm,clip]{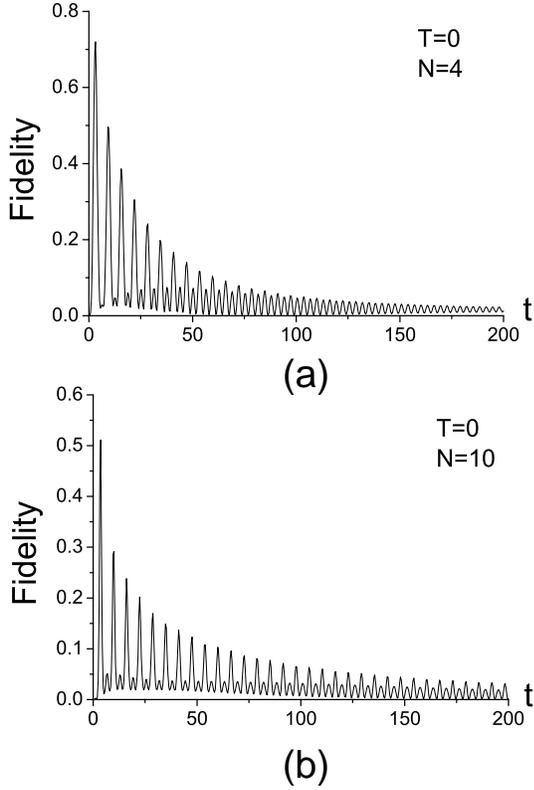}
\caption{\textit{The MTF as the function of time with the coupling strength
in Eq.\ref{couple}. The number of site in environment is assumed ten
times than that of chain, and parameter $\alpha=0.1\theta$ (a) The total
number $N$ of sites placed on the chain is $N=4$. (b) The total number $N$
of sites placed on the chain is $N=10$. The time is in unit of $1/\theta$.}}
\label{fN-4}
\end{figure}

Next we assume the coupling strength is of the form%
\begin{equation}
g_{xl}=\frac{\sqrt{\alpha }}{\pi ^{1/4}}\exp \left[ -\frac{\alpha ^{2}}{2}%
\left( l-x\right) ^{2}\right],
\label{couple}
\end{equation}
and the bath energy is a random Gaussian function of the site. To show the
bath effect on the QST in this case, in Fig.\ref{fN-4}, we numerically
plot the MTF as the function of time for $N=4$ and $N=10$ respectively.
In each figure, the number of environment site is assumed ten times
than that of chain; the parameter $\alpha=0.1\theta$.
Fig\ref{fN-4}(a) is plotted when the bath energy satisfy a normal
distribution (with respect to $x$) with mean $2.5\theta$ and
variance $\theta$; Fig\ref{fN-4}(b) is plotted when the normal
distribution of bath energy has the mean $5\theta$ and variance
$2.5\theta$. It shows that the transport efficiency decays rapidly
as time increase, and it also become worse as the the number of
sites increased.

\section{\label{sec:temper}effects on the QST at finite temperature}

In this section, we consider how the spatially distributed bath affects the
QST at finite temperature. Since the bath is initially in a thermal state
before coupling to the system, which is not a pure state. The
MTF can not completely reflect the transport efficiency, as it lacks the
information of state amplitude. Here, we employ the fidelity as the
criterion for transport efficiency. We consider the following situation: the
arbitrary state to be transferred is prepared in the $1$th site, and it is a
superposition of the single particle and the vacuum state.
\begin{equation}
\left\vert \psi _{1}\right\rangle =\alpha \left\vert 0_{1}\right\rangle
+\beta \left\vert 1_{1}\right\rangle ,
\end{equation}
where the footnote indicates the position of the site. The QST is completed
when arbitrary state is found in the end of chain. We assume that at $t=0$,
the chain is in a state
\begin{equation}
\left\vert \phi _{i}\right\rangle =\left\vert \psi \right\rangle
_{1}\left\vert \left\{ 0\right\} \right\rangle,
\end{equation}
with $\left\vert \psi \right\rangle $ being encoded in site one and no other
particle in chain. The fidelity is defined as%
\begin{equation}
F=Tr\left( U\left\vert \phi _{i}\right\rangle \left\langle \phi
_{i}\right\vert \rho _{B}U^{\dag }\left\vert \psi _{N}\right\rangle
\left\langle \psi _{N}\right\vert \right) ,  \label{definf}
\end{equation}
where $U(t)=\exp {(-i\tilde{H}t)}$ and $\rho _{B}=e^{-\beta H}/z$, and $z$
is the partition function. Eq. (\ref{definf}) can be rewritten as
\begin{equation}
F=Tr\left\langle \phi _{i}\right\vert \rho _{B}U^{\dag }\left\vert \psi
_{N}\right\rangle \left\langle \psi _{N}\right\vert U\left\vert \phi
_{i}\right\rangle.
\end{equation}
By defining sixteen components of fidelity
\begin{equation}
F_{ijlm}=tr\left[ U\left( t\right) \left\vert i_{1}\right\rangle
\left\langle j_{1}\right\vert \rho _{B}U^{\dag }\left( t\right) \left\vert
l_{N}\right\rangle \left\langle m_{N}\right\vert \right],
\end{equation}
where $i,j,l,m\in \left\{ 0,1\right\} $, the fidelity $F$ can be written as
the sum of the above sixteen fidelities. There are only ten terms
independent. In terms of operator $\hat{C}^{\dagger }$ and $\hat{C}$ defined in
section \ref{sec:Langevin}, all ten components of fidelity $F$ becomes
\begin{subequations}
\begin{eqnarray}
F_{0000} &=&\sum_{nm}d_{Nn}\left( \frac{\pi }{2}\right) d_{mN}\left( -\frac{%
\pi }{2}\right) \nonumber \\
&&\left\langle \left\{ 0\right\} \right\vert \left\langle \hat{C}_{n}\hat{C}%
_{m}^{\dag }\right\rangle _{B}\left\vert \left\{ 0\right\} \right\rangle
e^{i\theta \left( m-n\right) t}, \\
F_{0100} &=&\sum_{nmj}d_{Nn}\left( \frac{\pi }{2}\right) d_{mN}\left( -\frac{%
\pi }{2}\right) d_{1j}\left( \frac{\pi }{2}\right) \nonumber \\
&&\left\langle \left\{ 0\right\} \right\vert \hat{C}_{j0}\left\langle \hat{C}%
_{n}\hat{C}_{m}^{\dag }\right\rangle _{B}\left\vert \left\{ 0\right\}
\right\rangle e^{i\theta \left( m-n\right) t},
\end{eqnarray}
\begin{eqnarray}
F_{1100} &=&\sum_{nmjj^{\prime }}d_{Nn}\left( \frac{\pi }{2}\right)
d_{mN}\left( -\frac{\pi }{2}\right) d_{1j}\left( \frac{\pi }{2}\right)
d_{j^{\prime }1}\left( -\frac{\pi }{2}\right)  \nonumber \\
&&\left\langle \left\{ 0\right\} \right\vert \hat{C}_{j0}\left\langle \hat{C}
_{n}\hat{C}_{m}^{\dag }\right\rangle _{B}\hat{C}_{j^{\prime }0}^{\dag
}\left\vert \left\{ 0\right\} \right\rangle e^{i\theta \left( m-n\right) t}, \\
F_{0010} &=&\sum_{m}d_{mN}\left( -\frac{\pi }{2}\right) \left\langle \left\{
0\right\} \right\vert \left\langle \hat{C}_{m}^{\dag }\right\rangle
_{B}\left\vert \left\{ 0\right\} \right\rangle e^{i\theta mt}, \\
F_{0110} &=&\sum_{mj}d_{mN}\left( -\frac{\pi }{2}\right) d_{1j}\left( \frac{%
\pi }{2}\right) \nonumber  \\
&&\left\langle \left\{ 0\right\} \right\vert \hat{C}_{j0}\left\langle \hat{C}%
_{m}^{\dag }\right\rangle _{B}\left\vert \left\{ 0\right\} \right\rangle
e^{i\theta mt}, \\
F_{1010} &=&\sum_{mj}d_{mN}\left( -\frac{\pi }{2}\right) d_{j1}\left( -\frac{%
\pi }{2}\right)  \nonumber \\
&&\left\langle \left\{ 0\right\} \right\vert \left\langle \hat{C}_{m}^{\dag
}\right\rangle _{B}\hat{C}_{j0}^{\dag }\left\vert \left\{ 0\right\}
\right\rangle e^{i\theta mt}, \\
F_{1110} &=&\sum_{mjj^{\prime }}d_{mN}\left( -\frac{\pi }{2}\right)
d_{1j}\left( \frac{\pi }{2}\right) d_{j^{\prime }1}\left( -\frac{\pi }{2}%
\right)  \nonumber \\
&&\left\langle \left\{ 0\right\} \right\vert \hat{C}_{j0}\left\langle \hat{C}%
_{m}^{\dag }\right\rangle _{B}\hat{C}_{j^{\prime }0}^{\dag }\left\vert
\left\{ 0\right\} \right\rangle e^{i\theta mt}, \\
F_{0011} &=&\sum_{mn}d_{mN}\left( -\frac{\pi }{2}\right) d_{Nn}\left( \frac{%
\pi }{2}\right)  \nonumber \\
&&\left\langle \left\{ 0\right\} \right\vert \left\langle \hat{C}_{m}^{\dag }%
\hat{C}_{n}\right\rangle _{B}\left\vert \left\{ 0\right\} \right\rangle
e^{i\theta \left( m-n\right) t}, \\
F_{0111} &=&\sum_{mnj}d_{mN}\left( -\frac{\pi }{2}\right) d_{Nn}\left( \frac{%
\pi }{2}\right) d_{1j}\left( \frac{\pi }{2}\right) \nonumber  \\
&&\left\langle \left\{ 0\right\} \right\vert \hat{C}_{j0}\left\langle \hat{C}%
_{m}^{\dag }\hat{C}_{n}\right\rangle _{B}\left\vert \left\{ 0\right\}
\right\rangle e^{i\theta \left( m-n\right) t}, \\
F_{1111} &=&\sum_{mnjj^{\prime }}d_{mN}\left( -\frac{\pi }{2}\right)
d_{Nn}\left( \frac{\pi }{2}\right) d_{1j}\left( \frac{\pi }{2}\right)
d_{j^{\prime }1}\left( -\frac{\pi }{2}\right) \nonumber  \\
&&\left\langle \left\{ 0\right\} \right\vert \hat{C}_{j0}\left\langle \hat{C}%
_{m}^{\dag }\hat{C}_{n}\right\rangle _{B}\hat{C}_{j^{\prime }0}^{\dag
}\left\vert \left\{ 0\right\} \right\rangle e^{i\theta \left( m-n\right) t},
\end{eqnarray}
\end{subequations}
where $\hat{C}_{i0}(i=j,j^{\prime })$ are the initial operator in the Heisenberg
representation. Using Eq.(\ref{tt}), and neglecting the $\tilde{g}_{mx}^{\ast }%
\tilde{g}_{lx}$ term when $m\neq l$, we can obtain the expression of
fidelity for arbitrary state, and some of the off diagonal fidelity
vanish, i.e.
\begin{eqnarray}
F_{0100} &=&0,\ F_{0010}=0, \nonumber \\
F_{1001} &=&0,\ F_{1110}=0, \\
F_{0111} &=&0. \nonumber
\end{eqnarray}
A little reckoning shows that, like the MTF in zero temperature,
fidelity $F$ decays exponential with time. It also can be known
that, when the amplitudes of the transport state satisfy $\left\vert
\alpha \right\vert =\left\vert \beta \right\vert $, the fidelity $F$
is independent of temperature; when $\left\vert \alpha \right\vert
>\left\vert \beta \right\vert $, the fidelity $F$ is a decreasing of
temperature; when $\left\vert \alpha \right\vert <\left\vert\beta
\right\vert $, the fidelity $F$ increases with temperature. But as
time increases, the influence of temperature becomes small and
small.

\begin{figure}[htbp]
\includegraphics[bb=75 260 538 689, width=8cm,clip]{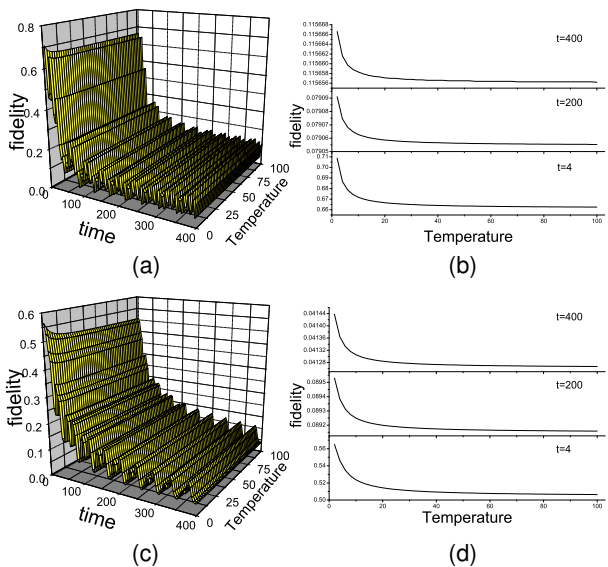}
\caption{\textit{Representation of fidelity $F$ as a function of time and
temperature with transport state $\left( \protect\sqrt{3}\left\vert
0_{1}\right\rangle +\left\vert 1_{1}\right\rangle \right) /2$. (a) 3-D
diagram for N=4. (b) the cross section for N=4. (c) 3-D diagram for N=10.
(d) the cross section for N=10. Other parameter are the same as Fig.\ref{fN-4}}}
\label{fN-4a10}
\end{figure}
\begin{figure}[htbp]
\includegraphics[bb=74 265 540 734, width=8cm,clip]{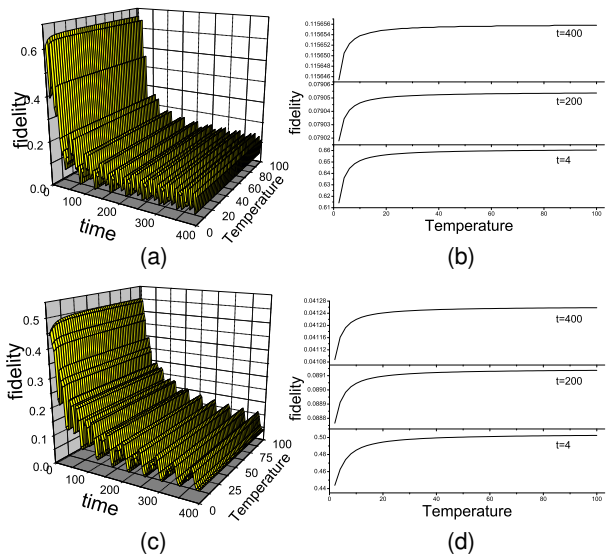}
\caption{\textit{Representation of fidelity $F$ as a function of time and
temperature with transport state $\left( \left\vert 0_{1}\right\rangle +%
\protect\sqrt{3}\left\vert 1_{1}\right\rangle \right) /2$.(a) 3-D diagram
for N=4. (b) the cross section for N=4. (c) 3-D diagram for N=10. (d) the
cross section for N=10. Other parameter are the same as Fig.\ref{fN-4}}}
\label{fN-10cs}
\end{figure}
To show the amplitude of the transport state influence, we depicted
the fidelity $F$ of different states as a function of time and temperature in
Fig.\ref{fN-4a10} and Fig.\ref{fN-10cs} respectively, in the case of coupling strength
in Eq.(\ref{couple}) and a random Gaussian distribution of the bath energy. Fig.
\ref{fN-4a10} is plotted when the transport state has the form $\left(\sqrt{3}
\left\vert 0_{1}\right\rangle+\left\vert 1_{1}\right\rangle\right)/2$. Fig.
\ref{fN-10cs} is plotted when the transport state is in $\left(\left\vert0_{1}
\right\rangle +\sqrt{3}\left\vert 1_{1}\right\rangle \right)/2$. Although the fidelity
is a decreasing function of time, it can be seen that, from the cross section (b) and (c)
in Fig.\ref{fN-4a10} and Fig.\ref{fN-10cs}, the amplitude value of transport state
determines the temperature effect on fidelity, that is, when the amplitude of state
$\left\vert0_{1}\right\rangle $ is larger than that of state $\left\vert
1_{1}\right\rangle $, the transport efficiency is a descending function of
temperature, and when the amplitude of state $\left\vert 0_{1}\right\rangle $
is smaller than that of state $\left\vert 1_{1}\right\rangle $, the
transport efficiency is a increasing function of temperature. From the cross
section (c) and (d) in Fig.\ref{fN-4a10} and Fig.\ref{fN-10cs}, It can be observed
that as time increase, the influence of temperature becomes small and small.
To show the influence of the site number $N$, in each figure, two 3D diagrams and its
cross sections are plotted for different site number $N$ of chain. It can be
obtained that the fidelity decreases as the number of sites increased. Hence
due to exchange of energy between chain and bath, we cannot obtain the original
state any longer as time increasing.

\section{\label{sec:end}Summary}

In summary, we have investigated the decoherence  problem in quantum
state transfer (QST). We show when an open spin chain of arbitrary
length N interacts with a globe distributed bath, the efficiency of
the QST decreases due to the exchange of energy between chain and
bath, but  the temperature effect on the QST is determined by the
form of transport state.

Although our discussion are based on the in the engineered tight
binding model with spinless Bloch electrons, it also can be applied
to an arbitrary many-particle state in such system. This is because
the irreducible tensor method in angular momentum theory could be
expected to work well for the QST in an engineered quantum models
with more electrons\cite{Chris1}. In the above studies, a qubit is
encoded in the superposition of the vacuum and the one-particle
state, which is  spinless  tight binding fermion. Actually, all the
conclusions we obtained can be extended to the spin electrons. The
electronic wave packet with spin polarization is an analogue of
photon ``flying qubit''. The spatial fidelity determines the
fidelity of freedom of spin.

This work is supported by the NSFC with grant Nos. 90203018, 10474104 and
60433050, and NFRPC with Nos. 2001CB309310 and 2005CB724508. The author Lan
Zhou gratefully acknowledges the support of K. C. Wong Education Foundation,
Hong Kong.

\end{document}